\documentstyle[preprint,aps]{revtex}

\newcommand{\be}{\begin{equation}}
\newcommand{\ee}{\end{equation}}
\newcommand{\bea}{\begin{eqnarray}}
\newcommand{\eea}{\end{eqnarray}}

\begin{document}
\preprint
\widetext

%%\include{psfig}
%\twocolumn[\hsize\textwidth\columnwidth\hsize\csname @twocolumnfalse\endcsname 
\title{SU(4) Theory for Spin Systems with Orbital
Degeneracy} \author{Y. Q. Li$^{1,2}$, Michael Ma$^1$, D. N. Shi$^{1,3}$, F.
C. Zhang$^1$}

\address{
$^1$ Department of Physics, University of Cincinnati, Cincinnati, Ohio
45221 \\
$^2$ Department of Physics, Zhejiang University, Hangzhou, China\\ $^3$
College of Science, Nanjing University of Aeronautics and Astronautics,
Nanjing, China\\
}
\date{\today}
\maketitle
\widetext
\begin{abstract}
\noindent
The isotropic limit of spin systems with orbital degeneracy is shown to
have global $SU(4)$ symmetry. On many 2D lattices, the ground state does
not posses long range order, which may explain the observed spin liquid
properties of $LiNiO_2$. In the $SU(4)$ Neel ordered state, spin-spin
correlations can be antiferromagneitc between two neighboring sites with
parallel magnetic moments. 

\end{abstract}


%\vskip2pc]

%\narrowtext

\newpage

In many transitional metal oxides, the electron configuration on the metal
ions has orbital degeneracy in addition to spin degeneracy. In these
systems, the sign and magnitude of the spin-spin couplings depends on the
orbital occupancy. This may result in interesting magnetic properties of
the Mott insulating phase, and is believed to be relevant to unusual
properties of many vanadium, titanium, manganese and nickel oxides
~\cite{McWhan,Cast,Word,Bao,Rice,Takano,Maekawa,Nagaosa,Shiba,Sawatzky,Zaanan,Kugel}. 
The Hamiltonian describing spin $s=1/2$ systems 
with a two-fold
orbital degeneracy (isospin $\tau =1/2$) was derived by Castellani 
${\it{et}\,{al.}}$~\cite{Cast}.
The Hamiltonian is rotationally symmetric in $\vec{s}$-space, but not in
$\vec{\tau}$-space. The aniostropy of the latter is due to Hund's rule and
the anisotropy in orbital wavefunctions.

In this Letter we study a simplified Hamiltonian, Eq. (1), where we focus
on the rotationally symmetric part in $\vec \tau$-space. The insight
learned from this higher symmetric model should shed light on our
understanding of more realistic systems.

We show the isotropic Hamiltonian (1) has a $SU(4)$ symmetry, and the spin
and orbital states are described by four flavors. The simplest $SU(4)$
singlet (flavorless) is a four-site cluster, in analogy to the $SU(2)$
singlet of a two-site pair. The model provides a new possibility for spin
liquid ground states in higher dimensions. For the square lattice, using
the fermion mean field theory, we find the flavor liquid state to be
stable against flavor or generalized spin density wave formation. By
comparing the energies of long-ranged ordered states to short-ranged ones
on the triangular lattice, we argue the ground state is likely to be a
resonant plaquette flavor liquid. In the $SU(4)$ Neel ordered state, the
spin-spin correlations can be antiferromagnetic (AF) between two
neighboring sites with {\it parallel} magnetic moments.

The simplest quantum spin-1/2 system with two-fold degenerate orbitals ($
\tau $-1/2) and rotational invariance in both $\vec s$ and $\vec \tau$
-spaces is given by ~\cite{Rice}

\begin{equation}
H=J\sum_{\langle i,j\rangle }(2\vec{s}_{i}\cdot
\vec{s}_{j}+1/2)(2\vec{\tau} _{i}\cdot \vec{\tau}_{j}+1/2), \label{S-T
Ham.} \end{equation}
where $\langle ij \rangle$ is the nearest neighbor (n.n.) pairs. For the AF coupling,
which we consider exclusively in this paper, $J>0$, and we set $J=1$
hereafter. Apparently, (\ref{S-T Ham.}) has $SU(2) \times SU(2)$ symmetry,
representing rotational invariance in both spin and orbital spaces, and also
interchange symmetry between spins and orbitals. We now show that the full
symmetry of (\ref{S-T Ham.}) is actually the higher symmery group $SU(4)$,
which unifies the spin and orbital degrees of freedom.

Let us start with an intuitive way to see the $SU(4)$ symmetry by rewriting
(\ref{S-T Ham.}) as,
\begin{equation}
H=(1/4)\sum_{\langle i,j\rangle }(\sum_{\gamma =1}^{15}A_{i}^{\gamma
}A_{j}^{\gamma }+1), \label{SU4 H}
\end{equation}
where $A^{\gamma }=2s^{\alpha },2\tau ^{\alpha },4s^{\alpha }\tau ^{\beta}$ 
for $\alpha ,\beta =x,y,z$. $A^{\gamma }$ can be considered as the
fifteen generators of the $SU(4)$ group. We now write them in terms of the
more standard generators  of group theory.

The Hamiltonian (\ref{S-T Ham.}) acts on a Hilbert space of $4$ basis
states at each site. Choosing these as $\mid s^{z},\tau ^{z}\rangle $, we
label them as

\begin{eqnarray}
&\mid &1\rangle =\mid 1/2,1/2\rangle ,\, \mid 2\rangle =\mid
-1/2,1/2\rangle , \nonumber \\
&\mid &3\rangle =\mid 1/2,-1/2\rangle ,\, \mid 4\rangle =\mid
-1/2,-1/2\rangle . \label{basis states}
\end{eqnarray}
These basis states form a fundamental representation of $SU(4)$. The
conventional $SU(4)$ generators $S_{m}^{n}$ acts on a basis state 
$\mid \mu \rangle$ according to
$S_{m}^{n}\mid \mu \rangle =\delta _{n,\mu }\mid m\rangle$. The
$S_{m}^{n}$'s are not hermitian, instead $(S_{m}^{n})^{\dag }=S_{n}^{m}$. Among
the fifteen generators $S_{m}^{n}$, twelve are non-diagonal with six
raising and six lowering operators. There is an identity 
$\sum_{m=1}^{4}S_{m}^{m}=1$, so that there are three independent diagonal
ones. The $S_{m}^{n}$'s are related to $\vec{s}$ and $\vec{\tau}$ by,
\begin{eqnarray}
2s^{z} &=&\sum_{m=1,3}(S_{m}^{m}-S_{m+1}^{m+1}),\hspace{0.2in}
s^{+}=\sum_{m=1,3}S_{m}^{m+1}, \nonumber \\ 2\tau ^{z}
&=&\sum_{m=1,2}(S_{m}^{m}-S_{m+2}^{m+2}),\hspace{0.2in}\tau
^{+}=\sum_{m=1,2}S_{m}^{m+2}, \label{spin-tau and S nm} 
\end{eqnarray}
where $s^{\pm }=s^{x}\pm is^{y}$, and 
$\tau ^{\pm }=\tau ^{x}\pm i\tau^{y}$.  The commutation relations for spin and orbital operators follow from the $SU(4)$ ones, 
$[S_{m}^{n},S_{k}^{l}]=\delta_{n,k}S_{m}^{l}-\delta _{m,l}S_{k}^{n}$. 
In terms of $S_{m}^{n}$, (\ref{S-T
Ham.}) becomes \begin{equation}
H=\sum_{\langle i,j\rangle }S_{m}^{n}(i)S_{n}^{m}(j), \label{SU4 H1}
\end{equation}
The repeated indices $n,m$ are summed in Eq. (\ref{SU4 H1}) and
hereafter. We conclude that $H$ has global $SU(4)$ invariance.

Eq. (\ref{S-T Ham.}), or equivalently (\ref{SU4 H1}), gives the effective Hamiltonian for the
corresponding Hubbard model in the large U-limit and at 1/4 filling,
\begin{equation}
H_{H}=-t\sum_{\langle i,j\rangle ,\mu }(c_{i,\mu }^{\dag },c_{j,\mu
}+h.c.)+U\sum_{i,\mu <\nu }n_{i,\mu }n_{i,\nu }, \label{Hubbard Ham}
\end{equation}
where $c_{i,\mu }$ and $c_{i,\mu }^{\dag }$ are annihilation and creation
operators of an electron at site $i$ and state $\mid \mu \rangle $.
In terms of electron operators, $S_{m}^{n}(i)=c_{i,m}^{\dag }c_{i,n}$.  
Eq. (\ref{Hubbard Ham}) is also equivalent to one of a class of models 
that has been solved by Sutherland in 1-dimension (1D) ~\cite{Sutherland}. 

We remark that the AF $SU(4)$ model here is different from the $SU(N)$
model studied by Sachdev and Read~\cite{Read}, and by
Arovas and Auerbach~\cite{Auerbach}. These authors considered AF $SU(N)$ model on 
bipartite lattices, where the two sublattices have conjugate representations
with respect to each other (''quarks'' and ''antiquarks''). 
In the present model, all the sites have 
the same representation, which is not self-conjugate.

To get insight on the physical properties,
we first consider systems with a few sites. Since $H$ has global $SU(4)$ invariance,
the eigenstates are given by irreducible representations of $SU(4)$. In
Fig. 1, we show the Young tableaux for two- and four-site systems. In the
two-site system, the lower energy ($\epsilon =-1$) states are 6-fold
degenerate (total spin $s=1$ and total orbital $\tau =0$ or $s=0$ and $\tau
=1$), and higher energy ($\epsilon =0$) states are 10-fold degenerate
($s=\tau =1$ or both $=0$). In the four-site system, the ground state is a
unique $SU(4)$ singlet $\mid SGL\rangle $, which is rotationally invariant 
under the SU(4) generators,

\begin{equation}
\sum_{i}A_{i}^{\gamma }\mid SGL\rangle =0,
\label{singlet}
\end{equation}
In terms of $S_m^n$, the singlet satisfies 
$\sum_{i}(S_m^n(i) -\delta _{mn}/4) \mid SGL \rangle =0$.  
A $SU(4)$ singlet is a singlet of spin, orbital, and the orbital-spin
crossing operator $U^{\alpha ,\beta }=4s^{\alpha }\tau ^{\beta }$, and is a
generalization of the $SU(2)$ singlet of spin only systems. The energy of
the $SU(4)$ singlet of the four-site is found to be $-N_{b}$, with $N_{b}$
the number of pairs $\langle ij \rangle$ in (\ref{S-T Ham.}). Hence, $N_{b}=4$ for a
4-site ring, $N_{b}=3$ for an open chain, and $N_{b}=6$ for a tetrahedron.
It is interesting to note that the energy of each bond in the four-site
system is $\epsilon_b =-1$, the best energy a single bond can have. This
would be difficult to understand from the conventional valence bond picture
for spin only systems, and again indicates the difference between (\ref{S-T
Ham.}) and spin only models including the four-site plaquette RVB state
recently discussed in literature~\cite{Lee}. 
In terms of the fermion operators of
(\ref{Hubbard Ham}), the $SU(4)$ singlet can be written as,

\begin{equation}
\mid SGL\rangle =\frac{1}{\sqrt{24}}\sum_{\{ijkl\}}c_{i1}^{\dag
}c_{j2}^{\dag }c_{k3}^{\dag }c_{l4}^{\dag }\mid 0\rangle , \label{singlet
fermion}
\end{equation}
where the sum is over all the permutations of the four sites ${ijkl}={1234}$.

For large number of sites, $N_{s}$,
the system can be a $SU(4)$ singlet
only if $N_{s}=4N$, with $N$ integers. This can be proved easily from
(\ref{singlet}). A necessary condition for a $SU(4)$ singlet is 
$\sum_{i}s_{i}^{z}=\sum_{i}\tau _{i}^{z}=\sum_{i}s_{i}^{z}\tau
_{i}^{z}=0$. This requires 1/4 of the sites of the system in each of the four
flavor states $\mid \mu \rangle $. Systems with $N_{s}\neq 4N$ may be
considered as ''edge state'' or excitations of $4N$ ones similar to
odd-number-site systems in the AF spin-1/2 systems. For an AF $SU(2)$ spin-1/2 chain, there is a theorem~\cite{Lieb} that 
the ground state is a
unique spin singlet for finite $N_{s}$ and, in the thermodynamic limit,
is either gapless or has broken translational symmetry (dimerization).
Affleck and Lieb generalized this to all $1/2$-integer $S$, and
to $SU(N)$\cite{Affleck1}. The $SU(N)$ representations for which their proof applies includes the fundamental
representation, which is the representaion for our problem. Thus, the ground state
of Eq. (\ref{S-T Ham.}) for a 1D chain is a unique $SU(4)$
singlet for finite $N_{s}=4N$,  and in the thermodynamic limit
is either gapless or breaks translational invariance
("quadmerization"). Their theorem for $SU(4)$ can be extended to 2D in
the same way as for the $SU(2)$ case, but requires a long narrow strip
as discussed by Affleck ~\cite{Affleck2}.
  
The $SU(4)$ symmetry identified for model (\ref{S-T Ham.}) has interesting
consequencies. Provided there is no symmetry breaking, it follows from the
symmetry that the thermodynamic correlation functions, denoted by $\langle
...\rangle $,

\begin{equation}
\langle s_{i}^{\alpha }s_{j}^{\alpha }\rangle =\langle \tau _{i}^{\alpha
}\tau _{j}^{\alpha }\rangle =\langle 4s_{i}^{\alpha }\tau _{i}^{\beta
}s_{j}^{\alpha }\tau _{j}^{\beta }\rangle =w_{ij}, \label{corr func sym}
\end{equation}
where $w_{ij}$ is a function of $i$ and $j$, and is independent of the
indices $\alpha $ and $\beta $ =$x,y,z$. This symmetry has been observed in
quantum Monte Carlo calculations of the 1D system~\cite{Beat}. For
translational invariant systems, the n.n. correlation is related to the
energy per bond, $\epsilon_{b}$,

\begin{equation}
w = \frac{1}{15}(\epsilon_b -\frac{1}{4}), \end{equation}

For a 1D chain, Sutherland's Bethe ansatz solution 
for the equivalent model~\cite{Sutherland} 
gives three gapless modes.  This may be interpreted in our model  
as pure spin, pure orbital, and orbital-spin crossing excitations.
The correlation
function $ w_{ij}$ is expected to decay in a power law at T=0,
and its sign
to be periodic with positive sign if $j-i=4N$, and negative otherwise. The
reason for the latter is the tendency for every four neighboring sites to
form a $SU(4)$ singlet.

The nature of the ground state of (\ref{S-T Ham.}) is of great potential
interest. There have been numerous theoretical activities since the discovery
of high temperature superconductivity to find possible spin liquid ground
states in two or three dimensions.  The additional orbital
degrees of freedom provides a new possibility for such states. Since there
are four $\it{equivalent}$ single site states $\mid \mu \rangle $ in (\ref{S-T
Ham.}) in comparison with two states in the spin only systems, we
expect quantum fluctuations to be stronger, making it more difficult to
establish long-range order, hence favoring flavor liquid states.

To illustrate this, we consider model (\ref{S-T Ham.}) on a square lattice
and carry out a fermion mean field theory. In fermion representation,
$H=-\sum_{\langle ij \rangle}\chi _{ij}^{\dag }\chi _{ij}+constant,$ where $\chi
_{ij}=\sum_{\mu =1}^{4}c_{i,\mu }^{\dag }c_{j,\mu }$. The model is similar
to the 2D $SU(N)$ t-J model of Affleck and Marston~\cite{Marston}, with the
important difference that here one fermion per site implies that
each flavor of fermions is $1/4$ filled, while in their study each flavor
of fermions is close to $1/2$ filled, the case relevant to cuprates. We
consider a uniform and real mean field bond amplitude 
$\chi =\langle \chi _{ij} \rangle$, and examine
its stability against a generalized spin density wave (SDW) state with four
sublattices $B_{\nu }, \nu=1,2,3,4$. The uniform mean field state describes a
flavor liquid. For the spin-1/2 Heisenberg model, the uniform state
was found unstable against the $(\pi ,\pi )$ SDW state~\cite{Yokoyama}. 
However, the instabilty is related to the nesting Fermi surface at $1/2$ filling. We expect the uniform
state to become stable against the SDW state at fillings sufficiently far away from 1/2, as in the present case. Minimizing the mean 
field energy with
respect to the SDW order parameter $m$, defined so that the mean occupation
number for the flavor $ \mu $ at site $i\subset B_{\nu }$, $\langle c_{i,\mu
}^{\dag }c_{i,\mu } \rangle =(1-m)/4+m\delta _{\nu ,\mu }$, we find that the
uniform bond state is stable against the SDW. Thus, the mean field theory
suggests the ground state of (\ref{S-T Ham.}) on the square lattice is
disordered. The flavor disordered state is found to be gapless in the
fermion mean field theory, but from previous studies of $SU(2)$ systems,
the issue of the gaplessness needs to be further examined. We note that 
the 1/4 filled uniform mean field state is unstable against the commensurate 
flux phase of flux $hc/(4e)$ per plaquette~\cite{Hasegawa,Lederer}. The 
flux phase is also a flavor liquid.

We expect the model on the triangular lattice to be also disordered. The AF
Heisenberg spin-1/2 system on the triangular lattice is believed to order
in a three-sublattice $120^{0}$ structure~\cite{Huse}. With orbital
degeneracy, such a spin ordering is no longer favored. In Fig. 2, we
compare the estimated energies for various long range ordered states,
including the classical $SU(4)$ Neel state (the same as the Ising like Neel
state for both triangular and square lattices, see
reference~\cite{Classical}), the orbital polarized spin ordered state, and
a valence bond state, with the $SU(4)$ singlet plaquette solid state. Unlike
the spin only problem, where the classical Neel state and the
valence bond solid state are degenerate, here the plaquette state has much
lower energy than all others. Since the plaquette state can resonate to
further lower the energy (and becomes a flavor liquid), we speculate the
ground state to be a resonant plaquette state with neither spin nor orbital
long range order. 

It is interesting to note that a spin liquid state with two-fold orbital
degeneracy may have already been realized in the best samples of
$LiNiO_{2}$ .  The measurements of magnetic
susceptibility, specific heat, $\mu $SR, and NMR at low temperature
show no long
range ordering in spins, and the $\mu $SR also shows that $Ni$-spins remain
fluctuating even at 20 mK~\cite{Takano}.
These evidence strongly indicate a spin liquid ground state. In that
material, a formal $Ni^{3+}$-ion has spin $s=1/2$ and a two-fold $e_{g}$
orbital degeneracy. The $Ni$-ions form layered triangular-lattices,
separated by two oxygen and one $Li$ layers, so that the interlayer
coupling is weak.  We
believe the orbital degeneracy is responsible for the observed spin liquid
properties, and Eq. (\ref{S-T Ham.}) may serve as a simple model to illustrate the role of the orbital degeneracy.

We now turn to the ordered states where the $SU(4)$ symmetry is broken.
There are many ways to break $SU(4)$, and here we discuss the generalized
4-sublattice Neel state shown in Fig. 2a. The remaining symmetry is 
$U(1) \times U(1) \times U(1)$. There are a total of 12
Goldstone modes, comprised of 4 spin-wave modes, 4 orbital-wave modes, and 4
spin-orbital-crossing modes.

The n.n. correlation function can have very unusual properties in states
with symmetry breaking. Consider for example the 4-sublattice Neel state. 
Let the two n.n. sites $i$ and $j$ belong to the two 
sublattices with $\langle s_{i}^{z}\rangle =\langle
s_{j}^{z}\rangle \geq 0$. In spin only systems, this would imply a
ferromagnetic coupling and the correlation $w(s^{z})\equiv \langle
s_{i}^{z}s_{j}^{z}\rangle >0$. In the presence of the orbital degeneracy,
the situation can be different. Let us start with the disordered state,
where $w(s^{z})<0$ from (10). Provided the transition is continuous, this implies that $w(s^{z})$ will remain negative at the transition point,
or close to it on the ordered side of the transition.  To estimate
the critical value of $\langle s_{i}^{z}\rangle $, we have carried out
calculations on a four-site system. We add a term in (1) representing
alternative fields coupled to the spin and orbital, so that the symmetry is
explicitly broken, and the four sites represent four sublattices. The
ground state and the correlation functions are calculated numerically. We
find $ w(s^{z})<0$ when $0<\langle s_{i}^{z}\rangle =\langle
s_{j}^{z}\rangle <0.18$. Therefore the system can have $\it{antiparallel}$
 spin correlation while both spins have $\it{parallel}$ magnetic moments. 
This unusual
magnetic property can be tested by neutron scattering.

We would like to thank I. Affleck, B. Frischmuth, N. Kawakami, B. Normand,
T. M. Rice, H. Shiba, M. Takano, K. Ueda for many stimulating and useful
discussions. FCZ wishes to thank Yukawa Institute, Kyoto University for its
hospitality during his visit, where part of this work was initiated. At the
completion of the work, we learned that K. Ueda ${\it {et}\,{al.}}$
have considered model (1) with emphasis in 1-dimension~\cite{Ueda}.

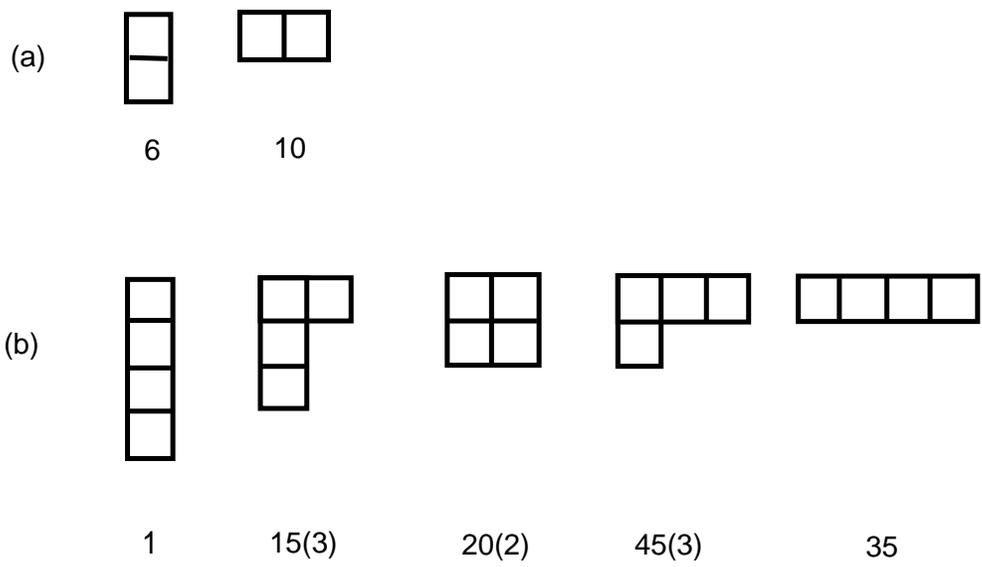
\begin{figure}
\caption{Young tableaux for $SU(4)$ model (1) a) in a two-site system, and
b) in a four-site system. The dimensionality of representations 
is indicated for each tableau, and inside the parenthesis is the number of distinct representation.}
\end{figure}

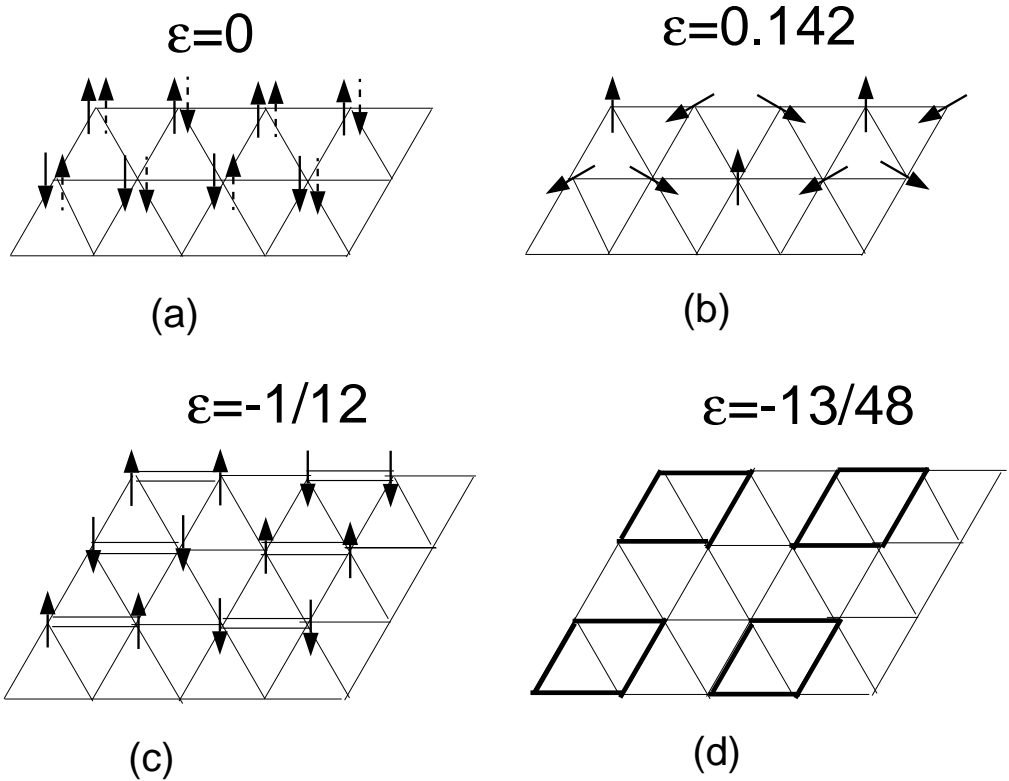
\begin{figure}
\caption{Average energy per bond $\epsilon$ of (1) for various states for
the triangular lattice. a) Classical 4-sublattice $SU(4)$ Neel state,
with each flavor shown by its orbital (dashed arrow) and spin (solid
arrow) state. b) State with orbital "ferromagnetic" and spin AF. 
In this case, Eq. (1) is reduced to 
$H=\sum_{\langle i,j \rangle}(2\vec{s}_{i}\cdot \vec{s}_{j}+1/2)$.
$\epsilon$ is deduced from Ref.~\protect\cite{Huse}. 
c) Valence bond
state. Each double-line represents a two-site valence bond of orbital
singlet and spin triplet. Note the spin long range order.
d) Plaquette state. Each plaquette (linked by four
thick lines) represents a four-site $SU(4)$ singlet.}
\end{figure}

\end{document}